\documentclass[subeqn,12pt,a4paper]{article}

\usepackage{amssymb,amsmath,amsfonts,amsthm, amscd, mathrsfs,helvet, mathtools}
\usepackage{bm, stmaryrd}
\usepackage{graphicx,verbatim}
\usepackage[usenames, dvipsnames]{color}
\usepackage{psfrag}
\usepackage[all]{xy}
\usepackage{cite}
\usepackage{srcltx}

\theoremstyle{plain}

\newtheorem*{theorem*}{Theorem}

\newtheorem*{proposition*}{Proposition}

\definecolor{brightBlue}{rgb}{0,0,1}

\definecolor{Violet}{rgb}{0.47,0,1}

\DeclareMathOperator{\tr}{tr}

\def\f{\mathfrak{f}}
\def\g{\mathfrak{g}}

\def\hfs{\widehat{\mathfrak{f}}^{\sigma}}

\def\ha{\mbox{\small $\frac{1}{2}$}}
\def\qa{\mbox{\small $\frac{1}{4}$}}

\def\A{\mathcal{A}}
\def\C{\mathcal{C}}

\def\L{\mathcal{L}}

\newcommand{\tensor}[1]{{\bf \underline{#1}}}

\def\1{\tensor{1}}
\def\2{\tensor{2}}
\def\3{\tensor{3}}
\def\4{\tensor{4}}

\def\beq{\begin{equation}}
\def\eeq{\end{equation}}
\def\beqz{\begin{equation*}}
\def\eeqz{\end{equation*}}
\def\bea{\begin{eqnarray}}
\def\eea{\end{eqnarray}}
\def\ads{{$AdS_5 \times S^5\,$}}
\numberwithin{equation}{section}
\def\Diag{{\mbox{diag}}}

\begin{document}

\begin{center}
\vspace*{2em}
{\large\bf
Generalized sine-Gordon models and
quantum braided groups}\\
\vspace{1.5em}
F. Delduc$\,{}^1$, M. Magro$\,{}^1$, B. Vicedo$\,{}^2$

\vspace{1em}
\begingroup\itshape
{\it 1) Laboratoire de Physique, ENS Lyon
et CNRS UMR 5672, Universit\'e de Lyon,}\\
{\it 46, all\'ee d'Italie, 69364 LYON Cedex 07, France}\\
\vspace{1em}
{\it 2) School of Physics, Astronomy and Mathematics,
University of Hertfordshire,}\\
{\it College Lane,
Hatfield AL10 9AB,
United Kingdom}
\par\endgroup
\vspace{1em}
\begingroup\ttfamily
Francois.Delduc@ens-lyon.fr, Marc.Magro@ens-lyon.fr, Benoit.Vicedo@gmail.com
\par\endgroup
\vspace{1.5em}
\end{center}

\paragraph{Abstract.}

We determine the quantized function algebras associated with various
examples of generalized sine-Gordon models.
These are quadratic algebras of the general Freidel-Maillet type, the classical
limits of which reproduce the lattice Poisson algebra recently obtained for these
models defined by a gauged Wess-Zumino-Witten action plus an integrable potential.
More specifically, we argue based on these examples that the natural framework
for constructing quantum lattice integrable versions of generalized sine-Gordon models
is that of affine quantum braided groups.

\section{Introduction}

The (semi-)symmetric space sine-Gordon models
constitute a broad class of generalizations of the sine-Gordon model. They may
be obtained through
Pohlmeyer reduction of (semi-)symmetric
space $\sigma$-models \cite{Pohlmeyer:1975nb,Grigoriev:2007bu}
(see \cite{Miramontes:2008wt} for a review).
Their Lagrangian formulation corresponds to (a fermionic extension of)
a gauged WZW model with an integrable potential
\cite{Bakas:1995bm,Grigoriev:2007bu}.
Like the sine-Gordon model itself, all these models are classically integrable.
However, a key difference
between them and the sine-Gordon model is that the Poisson algebra
satisfied by their Lax matrix is non-ultralocal. Yet a remarkable feature of
this particular non-ultralocal Poisson algebra, recently computed in
\cite{Delduc:2012qb,Delduc:2012mk,Delduc:2012vq}, is that it admits an
integrable lattice discretization. This promising result suggests that one may
be able to define quantum integrable lattice versions of generalized sine-Gordon
models.

This hope of being able to construct quantum lattice models for a
whole class of non-ultralocal integrable models is an entirely new prospect in
the study of non-ultralocality. As indicated above, a first step towards this
goal came from the determination of the lattice Poisson algebra for generalized
sine-Gordon models. In the present article we take a further step in this
direction by quantizing the lattice Poisson algebra obtained in
\cite{Delduc:2012qb,Delduc:2012mk,Delduc:2012vq}. More precisely, we determine
the quantized function algebra associated with different examples of generalized
sine-Gordon models.

Before indicating the plan of this article, let us recall that the simplest
generalization
of the sine-Gordon model, which is also taken as the first example in the
present work, corresponds to the complex sine-Gordon model
\cite{Lund:1976ze,Lund:1977dt,Getmanov:1977hk}. In the
continuum theory, the Poisson algebra satisfied by its Lax matrix was computed
in \cite{Maillet:1985ek}. It does not satisfy the criteria which enable the
construction of a corresponding lattice Poisson
algebra. However, as recalled above, the situation is quite different if one
views the complex sine-Gordon model as defined by a $SU(2)/U(1)$ gauged WZW
action plus an integrable potential. This is the standpoint taken in
this
article. Note that there are also indications \cite{Dorey:1994mg,Hoare:2010fb}
within factorized scattering theory that the proper definition of the quantum
complex sine-Gordon model is at the level of a gauged WZW model.

\medskip

The content of this article is divided in two parts. The first one, which
corresponds to sections \ref{sec1} and \ref{quantumlattice}, deals with general
results. Examples are
then presented in the second part, comprised of sections \ref{secsymsg} and
\ref{secgl44}.

The first part begins with a brief review of the results obtained in
\cite{Delduc:2012qb,Delduc:2012mk,Delduc:2012vq}. The Poisson algebra satisfied
by the continuum Lax matrix of (semi-)symmetric space sine-Gordon models is
recalled in section \ref{sec11}. The corresponding lattice Poisson algebra is
then given in section \ref{sec12}. It forms the
starting point of the analysis carried out in the rest of the article. This
lattice Poisson algebra is of the quadratic $abcd$-type
\cite{Freidel:1991jx,Freidel:1991jv} and depends on four matrices $a$,
$b$, $c$ and
$d$. These satisfy a number of properties which include those required to ensure
antisymmetry of the corresponding Poisson bracket, the Jacobi identity and
finally the existence of infinitely many commuting quantities.

The general analysis for the quantum case is performed in section
\ref{quantumlattice}.
To quantize the lattice Poisson algebra from section \ref{sec1}, we search for a
quantum lattice algebra of the general quadratic
$\mathcal{A}\mathcal{B}\mathcal{C}\mathcal{D}$-type
\cite{Freidel:1991jx,Freidel:1991jv}. As usual, the matrices $\mathcal{A}$,
$\mathcal{B}$, $\mathcal{C}$ and $\mathcal{D}$ should tend to the identity in
the classical limit $\hbar \to 0$ and reproduce the matrices $a$, $b$, $c$ and
$d$, respectively, at the next order. We give a list of natural conditions on
$\mathcal{A}$, $\mathcal{B}$, $\mathcal{C}$ and $\mathcal{D}$ which reduce in
the classical limit to those satisfied by $a$, $b$, $c$ and $d$ in section
\ref{sec12}. Among these are the conditions required in the general construction
of \cite{Freidel:1991jx,Freidel:1991jv}. Taken altogether, these properties lead
to the more refined structure of an affine quantum braided group
\cite{Hlavaty:1992pz,Hlavaty:1994rd}, as explained in section \ref{qbgsec}.

Concerning the second part, section \ref{secsymsg} is devoted to examples of
symmetric space sine-Gordon models. The first model considered is the complex
sine-Gordon model.
We then go on to consider models related to the affine Lie algebras $A_2^{(1)}$
and $A_2^{(2)}$. They correspond to the Pohlmeyer
reduction of the $\mathbb{C}P^2$ and $SU(3)/SO(3)$ $\sigma$-models,
respectively.
In section 5, we initiate the analysis for the \ads semi-symmetric space
sine-Gordon model \cite{Grigoriev:2007bu,Mikhailov:2007xr} by considering
the case of the twisted affine loop algebra of $\mathfrak{gl}(4|4)$.

\section{Quadratic Poisson algebra}
\label{sec1}

\subsection{Poisson algebra in the continuum}
\label{sec11}

As mentioned in the introduction, symmetric space sine-Gordon models are
obtained by  Pohlmeyer reduction of $\sigma$-models on symmetric spaces $F/G$.
We start this section by recalling the classical integrable structure of the
resulting gauged WZW models with an integrable potential.   We then indicate the
generalization to semi-symmetric space sine-Gordon models. This section is based
on the results in \cite{Delduc:2012qb,Delduc:2012mk,Delduc:2012vq}, to which the
reader is referred for more details.

Let $\f= \text{Lie}(F)$ be a Lie algebra equipped with a
$\mathbb{Z}_2$-automorphism
$\sigma : \f \to \f$, namely such that $\sigma^2 = \text{id}$, and let $\g =
\f^{(0)}= \{ x \in \f \,|\, \sigma(x) = x \}$ and $\f^{(1)}$ denote the eigenspaces of
$\sigma$ with eigenvalue $\pm 1$. The phase space of the theory
is parametrized by a pair of fields
$g$ and $A$ taking values in $G$ and $\g = \text{Lie}(G)$ respectively,
with Poisson brackets \cite{Bowcock:1988xr,Delduc:2012qb}
\begin{subequations} \label{PB}
\begin{align}
\label{PB-gg} \{ g_{\1}(\sigma), g_{\2}(\sigma') \} &= 0,\\
\label{PB-gA} \{ g_{\1}(\sigma), A_{\2}(\sigma') \} &= - 2 g_{\1}(\sigma)
C^{(00)}_{\1\2} \delta_{\sigma \sigma'},\\
\label{PB-AA} \{ A_{\1}(\sigma), A_{\2}(\sigma') \} &= - 2 \big[
C^{(00)}_{\1\2}, A_{\2}(\sigma) \big] \delta_{\sigma \sigma'} +
2 C^{(00)}_{\1\2} \partial_{\sigma} \delta_{\sigma \sigma'}.
\end{align}
\end{subequations}
 We denote by $C^{(00)} + C^{(11)}$ the decomposition of
the tensor Casimir $C$ of $\f$ with respect to the $\mathbb{Z}_2$-grading
induced by the involution $\sigma$.
The Lax matrix is given by
\begin{equation} \label{Pohl Lax cont}
\L(\sigma, \lambda) = A(\sigma) + \ha \lambda^{-1} \mu_-
g^{-1}(\sigma) T_- g(\sigma) - \ha \lambda \mu_+ T_+,
\end{equation}
where  $T_{\pm} \in \f^{(1)}$ and $\mu_{\pm} \in \mathbb{R}$ are
constants.
It takes values in the twisted polynomial loop algebra $\hfs
\subset \f [\lambda, \lambda^{-1}]$ of $\f $.
 The Lax matrix \eqref{Pohl Lax cont} satisfies the
non-ultralocal Poisson algebra
\begin{multline} \label{pbLcont}
\{ \L_{\1}(\sigma, \lambda), \L_{\2}(\sigma',\mu) \} = \big[
r_{\1\2}(\lambda/\mu),
\L_{\1}(\sigma,\lambda) + \L_{\2}(\sigma, \mu) \big] \delta_{\sigma \sigma'}\\
+ \big[ s_{\1\2}, \L_{\1}(\sigma, \lambda) - \L_{\2}(\sigma,\mu) \big]
\delta_{\sigma \sigma'} + 2 s_{\1\2} \partial_{\sigma} \delta_{\sigma \sigma'},
\end{multline}
where the  matrices $r$ and $s$ explicitly read
\begin{equation} \label{r-s matrices}
r_{\1\2}(\lambda) = \frac{1 + \lambda^2}{1 - \lambda^2}
C^{(00)}_{\1\2} + \frac{2 \lambda }{1 - \lambda^2} C^{(11)}_{\1\2},
\qquad s_{\1\2} = C^{(00)}_{\1\2}.
\end{equation}
The sum $r+s$ of the matrices in \eqref{r-s matrices} is a
non-skew-symmetric solution of the modified classical Yang-Baxter
equation (mCYBE) on $\hfs$, which underlies
the integrable structure of the model \cite{Vicedo:2010qd}.

In the case of a semi-symmetric space sine-Gordon model such as
the one associated with \ads \cite{Grigoriev:2007bu,Mikhailov:2007xr}, the
involutive automorphism $\sigma$ is replaced by a $\mathbb{Z}_4$-automorphism
with respect to which the Casimir decomposes as
$C = C^{(00)} + C^{(13)} + C^{(22)} + C^{(31)}$.
The Poisson brackets \eqref{PB} have to be supplemented with the
Poisson brackets of the fermionic fields. The corresponding Lax matrix,
whose expression may be found in \cite{Delduc:2012mk}, also satisfies
the algebra \eqref{pbLcont} but where now
\begin{equation} \label{r-sz4}
r_{\1\2}(\lambda) = \frac{1 + \lambda^4}{1 - \lambda^4}
C^{(00)}_{\1\2} + \frac{2 \lambda }{1 - \lambda^4} C^{(13)}_{\1\2}
+ \frac{2 \lambda^2 }{1 - \lambda^4} C^{(22)}_{\1\2}
+ \frac{2 \lambda^3 }{1 - \lambda^4} C^{(31)}_{\1\2},
\qquad s_{\1\2} = C^{(00)}_{\1\2}.
\end{equation}

The fact that the matrix $s_{\1\2}$ associated with the
(semi-)symmetric space sine-Gordon models is simply
the projection onto the subalgebra $\g$ of
constant loops in $\hfs$ is crucial. Indeed, it enables to define a
lattice discretization of the Poisson algebra \eqref{pbLcont}.
Furthermore, as we will see, this has important
consequences for the quantum case.

\subsection{Lattice Poisson algebra}
\label{sec12}

The lattice Poisson algebra
corresponding to \eqref{pbLcont} in the continuum limit is
\begin{subequations} \label{lattice algebra}
\begin{align}
\{ \L^n_{\1}(\lambda), \L^n_{\2}(\mu) \} &= a_{\1\2}(\lambda/\mu)
\L^n_{\1}(\lambda) \L^n_{\2}(\mu) -
 \L^n_{\1}(\lambda)
\L^n_{\2}(\mu) d_{\1\2}(\lambda/\mu) ,\\
\{ \L^n_{\1}(\lambda), \L^{n+1}_{\2}(\mu) \} &=- \L^{n+1}_{\2}(\mu)
c_{\1\2} \L^n_{\1}(\lambda), \label{lnnp1}\\
\{ \L^{n+1}_{\1}(\lambda), \L^n_{\2}(\mu) \} &=   \L^{n+1}_{\1}(\lambda)
 b_{\1\2} \L^n_{\2}(\mu), \label{lnp1n}\\
\{ \L^n_{\1}(\lambda), \L^m_{\2}(\mu) \} &= 0, \qquad |n - m| \geq 2,
\end{align}
\end{subequations}
where the lattice Lax matrix $\L^n$ encodes the physical degrees of freedom
at the $n^{\rm th}$ site of the lattice.
On the lattice, the property of non-ultralocality is encoded in
the Poisson brackets \eqref{lnnp1} and \eqref{lnp1n} which express the fact that
Lax matrices at adjacent sites $n$ and $n+1$ do not Poisson commute.
The Poisson algebra \eqref{lattice algebra} fits into the general scheme of
quadratic $abcd$-algebras considered in \cite{Freidel:1991jx,Freidel:1991jv}.
However, in the present case, the four matrices   $a$, $b$, $c$ and $d$ are
expressed
in terms of the  matrices  $r$ and $s$ given in \eqref{r-s matrices} or
\eqref{r-sz4}
together with \cite{SemenovTianShansky:1995ha} a skew-symmetric solution
$\alpha$ of the mCYBE on $\g$ as follows
\begin{equation} \label{defofabcd}
a(\lambda) = r(\lambda) + \alpha, \qquad b = -s - \alpha, \qquad c = -s +
\alpha, \qquad d(\lambda) = r(\lambda) - \alpha.
\end{equation}
In particular, $b$ and $c$ do not depend on the spectral parameter.
By virtue of their explicit expressions \eqref{defofabcd}, the matrices $a$,
$b$, $c$ and $d$ satisfy the following properties:

\begin{itemize}
  \item The first set of properties ensures that equations \eqref{lattice
algebra} define a Poisson bracket.  They are
\begin{equation} \label{skew prop abcd}
a_{\1\2}(\lambda) = - a_{\2\1}(\lambda^{-1}), \qquad d_{\1\2}(\lambda)
= - d_{\2\1}(\lambda^{-1}), \qquad b_{\1\2} = c_{\2\1}
\end{equation}
for the antisymmetry and
\begin{subequations} \label{condybabcd}
 \begin{align}
[a_{\1\2}(\lambda/ \mu),a_{\1\3}(\lambda)] + [a_{\1\2}(\lambda /\mu),
a_{\2\3}(\mu)] + [ a_{\1\3}(\lambda), a_{\2\3}(\mu)] &= 0,\\
[d_{\1\2}(\lambda/ \mu),d_{\1\3}(\lambda)] + [d_{\1\2}(\lambda /\mu),
d_{\2\3}(\mu)] + [ d_{\1\3}(\lambda), d_{\2\3}(\mu)] &= 0,\\
[a_{\1\2}(\lambda),c_{\1\3}] +[a_{\1\2}(\lambda),c_{\2\3}]+[c_{\1\3},c_{\2\3}]
&= 0,\\
[d_{\1\2}(\lambda),b_{\1\3}] +[d_{\1\2}(\lambda),b_{\2\3}]+[b_{\1\3},b_{\2\3}]
&= 0
\end{align}
\end{subequations}
for the Jacobi identity.

  \item An additional property of the matrices $b$ and $c$, which is not
  required in the general formalism of \cite{Freidel:1991jx,Freidel:1991jv},
is that they are themselves solutions of the classical Yang-Baxter equation
\begin{subequations} \label{cybeforbc}
\begin{align}
\label{cybeforb} [b_{\1\2}, b_{\1\3}] + [b_{\1\2}, b_{\2\3}] + [b_{\1\3},
b_{\2\3}] &= 0,\\
\label{cybeforc} [c_{\1\2}, c_{\1\3}] + [c_{\1\2}, c_{\2\3}] + [c_{\1\3},
c_{\2\3}] &= 0.
\end{align}
\end{subequations}
This is a consequence of the facts that $\alpha$ is a
solution of the mCYBE on $\g$ and that $s$ identifies with the Casimir on $\g$.

  \item Another important property which ensures that
the algebra \eqref{lattice algebra} leads to the existence of an infinite family
of commuting integrals of motion reads
\begin{equation} \label{classical integrability property}
a(\lambda) + b = c + d(\lambda).
\end{equation}
Indeed, introducing the monodromy $T = \L^N \dots \L^1$, its Poisson bracket can
be derived from the local lattice Poisson algebra \eqref{lattice algebra} using
the relation
\eqref{classical integrability property} and reads
\begin{align*}
\{ T_{\1}(\lambda), T_{\2}(\mu) \}  &= a_{\1\2}(\lambda/\mu) T_{\1}(\lambda)
T_{\2}(\mu) + T_{\1}(\lambda) b_{\1\2} T_{\2}(\mu)\\
&\qquad\qquad - T_{\2}(\mu) c_{\1\2} T_{\1}(\lambda) - T_{\2}(\mu)
T_{\1}(\lambda) d_{\1\2}(\lambda/\mu).
\end{align*}
It then immediately follows using \eqref{classical integrability property} once more
that the quantities $\tr\big( T^p(\lambda) \big)$ Poisson commute.

  \item Finally, the  matrices $r$ and $s$, in \eqref{r-s matrices}
as well as \eqref{r-sz4}, are related by
$\lim_{\lambda \to 0} r(\lambda) = s$ and $\lim_{\lambda \to \infty} r(\lambda)
= -s$.
An immediate consequence of this is that
\begin{subequations} \label{d limits}
\begin{alignat}{2}
\lim_{\lambda \to 0} a(\lambda) &= - b, &\qquad \qquad
\lim_{\lambda \to \infty} a(\lambda) &= c,\\
\lim_{\lambda \to 0} d(\lambda) &= - c, &\qquad \qquad
\lim_{\lambda \to \infty} d(\lambda) &= b.
\end{alignat}
\end{subequations}
\end{itemize}

\section{Quantum lattice algebra}
\label{quantumlattice}

\subsection{Quadratic algebra}
\label{subsec quant quad algebra}
On general grounds,
the quantum lattice algebra, whose classical limit corresponds to the Poisson
algebra \eqref{lattice algebra}, should be of the following form
\cite{Freidel:1991jx,Freidel:1991jv}
\begin{subequations} \label{quantum lattice algebra}
\begin{align}
\mathcal{A}_{\1\2}(q, \lambda/\mu) \hat{\L}^n_{\1}(\lambda) \hat{\L}^n_{\2}(\mu)
&= \hat{\L}^n_{\2}(\mu) \hat{\L}^n_{\1}(\lambda) \mathcal{D}_{\1\2}(q,
\lambda/\mu),\\
\hat{\L}^n_{\1}(\lambda) \hat{\L}^{n+1}_{\2}(\mu) &= \hat{\L}^{n+1}_{\2}(\mu)
\mathcal{C}_{\1\2}(q) \hat{\L}^n_{\1}(\lambda),\\
\hat{\L}^{n+1}_{\1}(\lambda) \mathcal{B}_{\1\2}(q) \hat{\L}^n_{\2}(\mu) &=
\hat{\L}^n_{\2}(\mu) \hat{\L}^{n+1}_{\1}(\lambda),\\
\hat{\L}^n_{\1}(\lambda) \hat{\L}^m_{\2}(\mu) &= \hat{\L}^m_{\2}(\mu)
\hat{\L}^n_{\1}(\lambda), \qquad |n - m| \geq 2,
\end{align}
\end{subequations}
where  $\hat{\L}^n = \L^n + O(\hbar)$ denotes the quantum lattice Lax matrix
which encodes the physical degrees of freedom at the $n^{\rm th}$ site of the
lattice. As usual, $q = e^{i \hbar}$ and the classical limit corresponds to
$\hbar \to 0$. In particular, one has in this limit
\begin{subequations} \label{ABCD classical limit}
\begin{gather}
\label{AB classical limit}
\mathcal{A}_{\1\2}(e^{i \hbar}, \lambda) = \mathsf{1} + i \hbar \,
a_{\1\2}(\lambda) + O(\hbar^2), \qquad
\mathcal{B}_{\1\2}(e^{i \hbar}) = \mathsf{1} + i \hbar \, b_{\1\2} +
O(\hbar^2),\\
\label{CD classical limit}
\mathcal{C}_{\1\2}(e^{i \hbar}) = \mathsf{1} + i \hbar \, c_{\1\2} + O(\hbar^2),
\qquad
\mathcal{D}_{\1\2}(e^{i \hbar}, \lambda) = \mathsf{1} + i \hbar \,
d_{\1\2}(\lambda) + O(\hbar^2).
\end{gather}
\end{subequations}
Besides having the correct classical limits, the quantum matrices $\mathcal{A}$,
$\mathcal{B}$, $\mathcal{C}$ and $\mathcal{D}$ satisfy certain further
properties which can be considered as the quantum analogs of those given in the
previous section for $a$, $b$, $c$ and $d$. Most of these properties ensure that
the algebra \eqref{quantum lattice algebra} is well defined and leads to the
existence of infinitely many commuting integrals of motion. The remaining
conditions are very natural from a mathematical point of view. The full list of
properties satisfied by the matrices $\mathcal{A}$, $\mathcal{B}$, $\mathcal{C}$
and $\mathcal{D}$ of sections \ref{secsymsg} and \ref{secgl44} is as follows:
\begin{itemize}
  \item  The first set of properties arises from considerations of the
consistency of the algebra \eqref{quantum lattice algebra}. By exchanging the
tensor indices $\mbox{\scriptsize $\1$} \leftrightarrow \mbox{\scriptsize $\2$}$
and the spectral parameters $\lambda \leftrightarrow \mu$ in equations
\eqref{quantum lattice algebra}, they may be rewritten as
\begin{align*}
\mathcal{A}_{\2\1}(q, \mu/\lambda)^{-1} \hat{\L}^n_{\1}(\lambda)
\hat{\L}^n_{\2}(\mu) &= \hat{\L}^n_{\2}(\mu) \hat{\L}^n_{\1}(\lambda)
\mathcal{D}_{\2\1}(q, \mu/\lambda)^{-1},\\
\hat{\L}^n_{\1}(\lambda) \hat{\L}^{n+1}_{\2}(\mu) &= \hat{\L}^{n+1}_{\2}(\mu)
\mathcal{B}_{\2\1}(q) \hat{\L}^n_{\1}(\lambda),\\
\hat{\L}^{n+1}_{\1}(\lambda) \mathcal{C}_{\2\1}(q) \hat{\L}^n_{\2}(\mu) &=
\hat{\L}^n_{\2}(\mu) \hat{\L}^{n+1}_{\1}(\lambda),\\
\hat{\L}^n_{\1}(\lambda) \hat{\L}^m_{\2}(\mu) &= \hat{\L}^m_{\2}(\mu)
\hat{\L}^n_{\1}(\lambda), \qquad |n - m| \geq 2.
\end{align*}
Therefore, to guarantee that these latter equations do not impose any new
relations on the quantum lattice Lax matrix $\hat{\L}^n$, we should require that
\begin{subequations}
\begin{align}
\mathcal{A}_{\1\2}(q, \lambda) \mathcal{A}_{\2\1}(q, \lambda^{-1}) &=
\mathcal{D}_{\1\2}(q, \lambda) \mathcal{D}_{\2\1}(q, \lambda^{-1}) \propto
\mathsf{1}, \label{unitAD} \\
\mathcal{C}_{\1\2}(q) &= \mathcal{B}_{\2\1}(q). \label{unitBC}
\end{align}
\end{subequations}
These are the quantum counterparts of the classical properties \eqref{skew prop
abcd}.

On the other hand, sufficient conditions for the consistency of the algebra \eqref{quantum
lattice algebra} read \cite{Freidel:1991jx,Freidel:1991jv}
\begin{subequations} \label{QYB relations}
\begin{align}
\label{AAA relation} \mathcal{A}_{\1\2}(q, \lambda/\mu) \mathcal{A}_{\1\3}(q,
\lambda) \mathcal{A}_{\2\3}(q, \mu) &= \mathcal{A}_{\2\3}(q, \mu)
\mathcal{A}_{\1\3}(q, \lambda) \mathcal{A}_{\1\2}(q, \lambda/\mu),\\
\label{DDD relation} \mathcal{D}_{\1\2}(q, \lambda/\mu) \mathcal{D}_{\1\3}(q,
\lambda) \mathcal{D}_{\2\3}(q, \mu) &= \mathcal{D}_{\2\3}(q, \mu)
\mathcal{D}_{\1\3}(q, \lambda) \mathcal{D}_{\1\2}(q, \lambda/\mu),\\\label{ACC
relation} \mathcal{A}_{\1\2}(q, \lambda) \mathcal{C}_{\1\3}(q)
\mathcal{C}_{\2\3}(q) &= \mathcal{C}_{\2\3}(q) \mathcal{C}_{\1\3}(q)
\mathcal{A}_{\1\2}(q, \lambda),\\
\label{DBB relation} \mathcal{D}_{\1\2}(q, \lambda) \mathcal{B}_{\1\3}(q)
\mathcal{B}_{\2\3}(q) &= \mathcal{B}_{\2\3}(q) \mathcal{B}_{\1\3}(q)
\mathcal{D}_{\1\2}(q, \lambda),
\end{align}
\end{subequations}
which constitute the quantum analogs of equations \eqref{condybabcd}.

  \item Since the classical matrices $b$ and $c$ satisfy the
CYBE \eqref{cybeforbc}, it is natural to seek matrices $\mathcal{B}(q)$ and
$\mathcal{C}(q)$ which are themselves solutions of the quantum Yang-Baxter
equation (QYBE). We shall therefore impose the following further conditions on
these matrices
\begin{subequations} \label{QYBE for B C}
\begin{align}
\label{BBB relation}
\mathcal{B}_{\1\2}(q) \mathcal{B}_{\1\3}(q) \mathcal{B}_{\2\3}(q) &=
\mathcal{B}_{\2\3}(q) \mathcal{B}_{\1\3}(q) \mathcal{B}_{\1\2}(q),\\
\label{CCC relation}
\mathcal{C}_{\1\2}(q) \mathcal{C}_{\1\3}(q) \mathcal{C}_{\2\3}(q) &=
\mathcal{C}_{\2\3}(q) \mathcal{C}_{\1\3}(q) \mathcal{C}_{\1\2}(q).
\end{align}
\end{subequations}
Although these properties are   not required in the general formalism of
\cite{Freidel:1991jx,Freidel:1991jv} for quadratic quantum lattice algebras of
the type \eqref{quantum lattice algebra}, they will play a very important role
for us in underpinning the algebraic structure underlying the integrable models
considered.

  \item Another property which  plays a central role in the interpretation of
the algebra \eqref{quantum lattice algebra}, to be described shortly, and which
we shall require
      our set of four quantum $R$-matrices $\mathcal{A}(q, \lambda)$,
$\mathcal{B}(q)$,  $\mathcal{C}(q)$ and $\mathcal{D}(q, \lambda)$ to satisfy is
\begin{equation} \label{AB=CD}
\mathcal{A}(q, \lambda) \mathcal{B}(q) = \mathcal{C}(q) \mathcal{D}(q, \lambda).
\end{equation}
Even though the classical limit of this equation is equivalent to the classical
property
\eqref{classical integrability property}, it is not the appropriate quantum
generalization of the latter.

Instead, the correct quantum analog of \eqref{classical integrability property}
is the existence of a numerical matrix $\gamma(q)$ satisfying the following relation
\begin{equation} \label{local to global}
\gamma_{\2}(q) \mathcal{B}_{\1\2}(q) \gamma_{\1}(q) \mathcal{A}_{\1\2}(q,
\lambda) = \mathcal{D}_{\1\2}(q, \lambda) \gamma_{\1}(q) \mathcal{C}_{\1\2}(q)
\gamma_{\2}(q).
\end{equation}
In order for this equation to reduce to \eqref{classical integrability property}
in the classical limit, the matrix $\gamma(q)$ should be such that it tends to
the identity matrix as $q \to 1$.
The property \eqref{local to global} is essential to ensure the
passage from the local commutation relations \eqref{quantum lattice algebra} to
the
global commutation relation \cite{Freidel:1991jx,Freidel:1991jv}
\begin{equation} \label{quantum monodromy algebra}
\mathcal{A}_{\1\2}(q, \lambda/\mu) \hat{T}_{\1}(\lambda) \mathcal{B}_{\1\2}(q)
\hat{T}_{\2}(\mu) = \hat{T}_{\2}(\mu) \mathcal{C}_{\1\2}(q)
\hat{T}_{\1}(\lambda) \mathcal{D}_{\1\2}(q, \lambda/\mu)
\end{equation}
for the quantum monodromy defined as $\hat{T}(\lambda) = \hat{\L}^N(\lambda)
\gamma(q) \hat{\L}^{N-1}(\lambda) \gamma(q) \ldots \gamma(q)
\hat{\L}^1(\lambda)$.
It is in this sense that the relation \eqref{local to global} is the quantum
analog of the classical property \eqref{classical integrability property}.

With the monodromy matrix so defined and satisfying the quadratic algebra
\eqref{quantum monodromy algebra}, the property which ultimately 
guarantees the
existence of an infinite
family of commuting operators is the existence of another numerical matrix
$\widetilde{\gamma}(q)$ such that
\begin{equation} \label{quantum integrability property}
\widetilde{\mathcal{A}}_{\1\2}(q, \lambda) \widetilde{\gamma}_{\1}(q)
\widetilde{\mathcal{B}}_{\1\2}(q) \widetilde{\gamma}_{\2}(q)
= \widetilde{\gamma}_{\2}(q) \widetilde{\mathcal{C}}_{\1\2}(q)
\widetilde{\gamma}_{\1}(q)
\widetilde{\mathcal{D}}_{\1\2}(q, \lambda)
\end{equation}
where the matrices $\widetilde{\mathcal{A}}$, $\widetilde{\mathcal{B}}$,
$\widetilde{\mathcal{C}}$ and $\widetilde{\mathcal{D}}$ are defined as
\begin{equation*}
\widetilde{\mathcal{A}} = (\mathcal{A}^{t_1t_2})^{-1}, \quad
\widetilde{\mathcal{B}}
= [ (\mathcal{B}^{t_1})^{-1}]^{t_2}, \quad \widetilde{\mathcal{C}} =
[ (\mathcal{C}^{t_2})^{-1}]^{t_1},
\quad \widetilde{\mathcal{D}} = (\mathcal{D}^{t_1t_2})^{-1}. \label{tildega}
\end{equation*}
Here $x^t$ denotes the (super-)transpose of $x$. In every example considered in
this article, $\widetilde{\gamma}(q)$ is a diagonal matrix tending to the
identity in the limit $q \to 1$ and is therefore also consistent in the
classical limit with the relation \eqref{classical integrability property}.

The global monodromy algebra \eqref{quantum monodromy algebra} together
with the property \eqref{quantum integrability property} ensure
\cite{Freidel:1991jx,Freidel:1991jv} that the operators
$\tr\big(\widetilde{\gamma}(q)^t\, \hat{T}(\lambda) \big)$ commute for
different values of the spectral parameter.

   \item In each example we also have the following relations
\begin{alignat*}{2}
\lim_{\lambda \to 0} \mathcal{A}(q, \lambda) &= \mathcal{B}(q)^{-1},
&\qquad \qquad
\lim_{\lambda \to \infty} \mathcal{A}(q, \lambda) &= \mathcal{C}(q),\\
\lim_{\lambda \to 0} \mathcal{D}(q, \lambda) &= \mathcal{C}(q)^{-1},
&\qquad \qquad
\lim_{\lambda \to \infty} \mathcal{D}(q, \lambda) &= \mathcal{B}(q)
\end{alignat*}
which are natural quantum analogs of \eqref{d limits}. Using these relations we
observe that \eqref{ACC relation}, \eqref{DBB relation} and \eqref{QYBE for B C}
can all be obtained as   appropriate  limits of the QYBE \eqref{AAA relation}
and \eqref{DDD relation}.
\end{itemize}

\subsection{Affine quantum braided group}
\label{qbgsec}

Given a set of four matrices $\mathcal{A}$, $\mathcal{B}$, $\mathcal{C}$ and
$\mathcal{D}$ constructed to satisfy all of the above properties, it turns out
that the algebraic structure underlying the quantum integrability of the
corresponding quantum model is precisely that of an affine  quantum braided
group \cite{Hlavaty:1992pz,Hlavaty:1994rd}.

Indeed, using the relation \eqref{AB=CD} together with \eqref{unitBC}, the four
matrices $\mathcal{A}$, $\mathcal{B}$, $\mathcal{C}$ and $\mathcal{D}$ may be
expressed in terms of just two matrices $\mathcal{R}$ and $\mathcal{Z}$ as
follows
\begin{alignat*}{2}
\mathcal{A}_{\1\2}(q, \lambda) &= \mathcal{Z}_{\2\1}(q) \mathcal{R}_{\1\2}(q,
\lambda) \mathcal{Z}_{\1\2}(q)^{-1}, &\qquad
\mathcal{B}_{\1\2}(q) &= \mathcal{Z}_{\1\2}(q),\\
\mathcal{D}_{\1\2}(q, \lambda) &= \mathcal{R}_{\1\2}(q, \lambda), &\qquad
\mathcal{C}_{\1\2}(q) &= \mathcal{Z}_{\2\1}(q),
\end{alignat*}
The relations in \eqref{unitAD} then translate into the single unitarity
condition
\begin{equation} \label{unitarity R}
\mathcal{R}_{\1\2}(q, \lambda) \mathcal{R}_{\2\1}(q, \lambda^{-1}) \propto
\mathsf{1}.
\end{equation}
Note that there is no unitarity condition on the matrix $\mathcal{Z}$.
Moreover, the full set of QYB-type relations \eqref{QYB relations} and
\eqref{QYBE for B C} is  equivalent to
\begin{subequations} \label{QYB relations RZ}
\begin{align}
\label{RRR relation} \mathcal{R}_{\1\2}(q, \lambda/\mu) \mathcal{R}_{\1\3}(q,
\lambda) \mathcal{R}_{\2\3}(q, \mu) &= \mathcal{R}_{\2\3}(q, \mu)
\mathcal{R}_{\1\3}(q, \lambda) \mathcal{R}_{\1\2}(q, \lambda/\mu),\\
\label{ZZZ relation} \mathcal{Z}_{\1\2}(q) \mathcal{Z}_{\1\3}(q)
\mathcal{Z}_{\2\3}(q) &= \mathcal{Z}_{\2\3}(q) \mathcal{Z}_{\1\3}(q)
\mathcal{Z}_{\1\2}(q),\\
\label{RZZ relation} \mathcal{R}_{\1\2}(q, \lambda) \mathcal{Z}_{\1\3}(q)
\mathcal{Z}_{\2\3}(q) &= \mathcal{Z}_{\2\3}(q) \mathcal{Z}_{\1\3}(q)
\mathcal{R}_{\1\2}(q, \lambda),\\
\label{ZZR relation} \mathcal{Z}_{\1\2}(q) \mathcal{Z}_{\1\3}(q)
\mathcal{R}_{\2\3}(q, \lambda) &= \mathcal{R}_{\2\3}(q, \lambda)
\mathcal{Z}_{\1\3}(q) \mathcal{Z}_{\1\2}(q).
\end{align}
\end{subequations}

In terms of the matrices $\mathcal{R}$ and $\mathcal{Z}$, the quantum monodromy
algebra \eqref{quantum monodromy algebra} then reads
\begin{equation} \label{QBG}
\mathcal{R}_{\1\2}(q, \lambda/\mu) \mathcal{Z}_{\1\2}(q)^{-1}
\hat{T}_{\1}(\lambda) \mathcal{Z}_{\1\2}(q) \hat{T}_{\2}(\mu) =
\mathcal{Z}_{\2\1}(q)^{-1} \hat{T}_{\2}(\mu) \mathcal{Z}_{\2\1}(q)
\hat{T}_{\1}(\lambda) \mathcal{R}_{\1\2}(q, \lambda/\mu),
\end{equation}
which is exactly the relation defining an affine quantum braided group as
introduced in \cite{Hlavaty:1992pz,Hlavaty:1994rd}.

In the remaining sections we present the quantum $R$-matrices $\mathcal{A}(q,
\lambda)$, $\mathcal{B}(q)$, $\mathcal{C}(q)$ and $\mathcal{D}(q, \lambda)$
entering the affine quantum braided group \eqref{QBG} for various models.

\section{Symmetric space sine-Gordon models}
\label{secsymsg}

\subsection{Complex sine-Gordon model}
\label{secCsG}

\paragraph{Automorphism.}
In the setup of section \ref{sec1}, consider the case of the Lie algebra
$\f = \mathfrak{su}(2) \oplus \mathfrak{su}(2)$ and
define the $\mathbb{Z}_2$-automorphism $\sigma : \f \to \f$ as the flip
\begin{equation*}
\sigma(x, y) = (y, x),
\end{equation*}
for any $x, y \in \mathfrak{su}(2)$. The corresponding eigenspaces of $\sigma$
read
\begin{equation} \label{f0 f1}
\g = \f^{(0)} = \{ (x, x) \,|\, x \in \mathfrak{su}(2) \}, \qquad
\f^{(1)} = \{ (x, -x) \,|\, x \in \mathfrak{su}(2) \}.
\end{equation}
Now introduce the standard basis for $\mathfrak{su}(2)$, namely
\begin{equation} \label{sl2 basis}
\mathsf{H} = \left( \begin{array}{cc} 1 & 0\\ 0 & -1 \end{array} \right), \qquad
\mathsf{E} = \left( \begin{array}{cc} 0 & 1\\ 0 & 0 \end{array} \right), \qquad
\mathsf{F}  = \left( \begin{array}{cc} 0 & 0\\ 1 & 0 \end{array} \right),
\end{equation}
in terms of which a basis of $\mathfrak{su}(2) \oplus \mathfrak{su}(2)$ reads
\begin{equation*}
\mathsf{H}_1 = (\mathsf{H}, 0), \quad
\mathsf{E}_1 = (\mathsf{E}, 0), \quad
\mathsf{F}_1 = (\mathsf{F}, 0), \qquad
\mathsf{H}_2 = (0, \mathsf{H}), \quad
\mathsf{E}_2 = (0, \mathsf{E}), \quad
\mathsf{F}_2 = (0, \mathsf{F}).
\end{equation*}
Let us also introduce the block matrices $\mathsf{1}_1 = (\mathsf{1}, 0)$ and
$\mathsf{1}_2 = (0, \mathsf{1})$  where $\mathsf{1}$ is the $2 \times 2$
identity matrix.
In terms of the above we may write down a basis of the subspaces
$\f^{(0)}$ and $\f^{(1)}$ as
\begin{equation} \label{graded basis}
\begin{split}
\mathsf{h}^{(0)} = \mathsf{H}_1 + \mathsf{H}_2, \qquad
\mathsf{e}^{(0)} &= \mathsf{E}_1 + \mathsf{E}_2, \qquad
\mathsf{f}^{(0)} = \mathsf{F}_1 + \mathsf{F}_2, \\
\mathsf{h}^{(1)} = \mathsf{H}_1 - \mathsf{H}_2, \qquad
\mathsf{e}^{(1)} &= \mathsf{E}_1 - \mathsf{E}_2, \qquad
\mathsf{f}^{(1)} = \mathsf{F}_1 - \mathsf{F}_2
\end{split}
\end{equation}
respectively.
\paragraph{Casimir decomposition.}
The tensor Casimir is the sum of the tensor Casimirs for each
$\mathfrak{su}(2)$, namely
\begin{equation*}
C = \ha \mathsf{H}_1 \otimes \mathsf{H}_1 + \mathsf{E}_1 \otimes \mathsf{F}_1 +
\mathsf{F}_1 \otimes \mathsf{E}_1 + \ha \mathsf{H}_2 \otimes \mathsf{H}_2 +
\mathsf{E}_2 \otimes \mathsf{F}_2 + \mathsf{F}_2 \otimes \mathsf{E}_2,
\end{equation*}
which can be decomposed as $C = C^{(00)} + C^{(11)}$ relative to $\f = \f^{(0)}
\oplus \f^{(1)}$, where
\begin{align*}
C^{(00)} &= \qa \mathsf{h}^{(0)} \otimes \mathsf{h}^{(0)} + \ha \mathsf{e}^{(0)}
\otimes \mathsf{f}^{(0)} + \ha \mathsf{f}^{(0)} \otimes \mathsf{e}^{(0)},\\
C^{(11)} &= \qa \mathsf{h}^{(1)} \otimes \mathsf{h}^{(1)} + \ha \mathsf{e}^{(1)}
\otimes \mathsf{f}^{(1)} + \ha \mathsf{f}^{(1)} \otimes \mathsf{e}^{(1)}.
\end{align*}

\paragraph{Classical $r$-matrices.}
We let the matrix $\alpha$ appearing in \eqref{defofabcd} be the standard
skew-symmetric solution of the mCYBE on $\g = \mathfrak{su}(2)$, namely
\begin{equation*}
\alpha = \ha \big( \mathsf{e}^{(0)} \otimes \mathsf{f}^{(0)} - \mathsf{f}^{(0)}
\otimes \mathsf{e}^{(0)} \big).\end{equation*}
The corresponding non-skew-symmetric solutions
$b$ and $c$ of the CYBE  defined by \eqref{defofabcd}
then read
\begin{equation} \label{alpha pm}
b = - \qa \mathsf{h}^{(0)} \otimes \mathsf{h}^{(0)} - \mathsf{e}^{(0)} \otimes
\mathsf{f}^{(0)}, \qquad
c = - \qa \mathsf{h}^{(0)} \otimes \mathsf{h}^{(0)} - \mathsf{f}^{(0)} \otimes
\mathsf{e}^{(0)}.
\end{equation}

In terms of these, the   spectral parameter dependent classical $r$-matrices in
\eqref{defofabcd} may be written as follows
\begin{equation*}
a(\lambda) = - \delta(\lambda) b + \big( \mathsf{1} - \delta(\lambda) \big) c,
\qquad
d(\lambda) = - \delta(\lambda) c + \big( \mathsf{1} - \delta(\lambda) \big) b,
\end{equation*}
where we have introduced the diagonal matrix
\begin{equation*}
\delta(\lambda) = \frac{1}{1 - \lambda} \big( \mathsf{1}_1 \otimes \mathsf{1}_1
+ \mathsf{1}_2 \otimes \mathsf{1}_2 \big) + \frac{1}{1 + \lambda} \big(
\mathsf{1}_1 \otimes \mathsf{1}_2 + \mathsf{1}_2 \otimes \mathsf{1}_1 \big).
\end{equation*}

\paragraph{Quantum $R$-matrices.}
Next we give quantizations of the above
classical $r$-matrices $a(\lambda)$, $b$, $c$ and $d(\lambda)$. Specifically,
these are solutions  $\mathcal{A}(q, \lambda)$,
$\mathcal{B}(q)$, $\mathcal{C}(q)$ and $\mathcal{D}(q, \lambda)$ of the QYBE
with
classical limits \eqref{ABCD classical limit} for $q = e^{i \hbar}$.
Quantizations
of the constant classical $r$-matrices $b$ and $c$ read
\begin{subequations} \label{quantum B C}
\begin{align}
\label{quantum B} \mathcal{B}(q) &= q^{- \frac{1}{4} \mathsf{h}^{(0)} \otimes
\mathsf{h}^{(0)}} + q^{-\frac{1}{4}} (1 - q) \, \mathsf{e}^{(0)} \otimes
\mathsf{f}^{(0)}, \\
\label{quantum C} \mathcal{C}(q) &= q^{- \frac{1}{4} \mathsf{h}^{(0)} \otimes
\mathsf{h}^{(0)}} + q^{-\frac{1}{4}} (1 - q) \, \mathsf{f}^{(0)} \otimes
\mathsf{e}^{(0)}.
\end{align}
\end{subequations}

In terms of these,  quantizations of $a(\lambda)$ and $d(\lambda)$ are
respectively given by
\begin{subequations} \label{quantum A D}
\begin{align}
\mathcal{A}(q, \lambda) &= \delta(q, \lambda) \mathcal{B}(q)^{-1} + \big(
\mathsf{1} - \delta(q, \lambda) \big) \mathcal{C}(q), \label{quantum A}\\
\!\!\! \mathcal{D}(q, \lambda) &= \delta(q, \lambda) \mathcal{C}(q)^{-1} + \big(
\mathsf{1} - \delta(q, \lambda) \big) \mathcal{B}(q), \label{quantum D}
\end{align}
\end{subequations}
where we have introduced the following $q$-deformation of the diagonal matrix
$\delta(\lambda)$,
\begin{equation*}
\delta(q, \lambda) = \frac{\sqrt{q}}{\sqrt{q} - \lambda} \big( \mathsf{1}_1
\otimes \mathsf{1}_1 + \mathsf{1}_2 \otimes \mathsf{1}_2 \big) + \frac{q}{q +
\lambda} \mathsf{1}_1 \otimes \mathsf{1}_2 + \frac{1}{1 + \lambda} \mathsf{1}_2
\otimes \mathsf{1}_1.
\end{equation*}

\paragraph{Properties.} One can check that  the quantum matrices in
\eqref{quantum B C} and \eqref{quantum A D} satisfy all the properties listed in
section \ref{quantumlattice}. In particular, relations  \eqref{local to global}
and \eqref{quantum integrability property} hold with   diagonal matrices
$\gamma(q)=\mathsf{1}$ and $
\widetilde{\gamma}(q) =   \Diag(q,1,q,1)$.

To describe the property \eqref{unitAD} of unitarity, consider the diagonal
matrix
\begin{equation*}
K(q, \lambda) = \delta(q, \lambda) \delta(q, q \lambda) \delta\big(q,
q^{-\frac{1}{2}} \lambda\big)^{-1} \delta\big(q, q^{\frac{3}{2}}
\lambda\big)^{-1}.
\end{equation*}
It commutes with all four $R$-matrices $\mathcal{A}(q, \lambda)$,
$\mathcal{B}(q)$, $\mathcal{C}(q)$ and $\mathcal{D}(q, \lambda)$ and tends to
the identity in the limits $\lambda \to 0$ and $\lambda \to \infty$.
Furthermore, in the limit
$\hbar \to 0$, one has
$K(e^{i \hbar}, \lambda) = \mathsf{1} + O(\hbar^2)$.  Therefore the rescaled
matrices $\widehat{\mathcal{A}}(q, \lambda) = K(q, \lambda)^{-\frac{1}{2}}
\mathcal{A}(q, \lambda)$ and $\widehat{\mathcal{D}}(q, \lambda) = K(q,
\lambda)^{-\frac{1}{2}} \mathcal{D}(q, \lambda)$ together with $\mathcal{B}(q)$
and $\mathcal{C}(q)$ satisfy all properties of section \ref{quantumlattice}
including the unitarity conditions
\begin{equation*}
\widehat{\mathcal{A}}_{\1\2}(q, \lambda) \widehat{\mathcal{A}}_{\2\1}(q,
\lambda^{-1}) = \widehat{\mathcal{D}}_{\1\2}(q, \lambda)
\widehat{\mathcal{D}}_{\2\1}(q, \lambda^{-1}) = \mathsf{1}.
\end{equation*}

\subsection{Models related to affine Lie algebras $A^{(n)}_2$}
\label{secA2}

In this section we consider generalized sine-Gordon models
associated with both the untwisted and twisted affine Lie algebras $A^{(1)}_2$
and $A^{(2)}_2$.

\subsubsection{$\mathbb{C}P^2$ symmetric space sine-Gordon model}
\label{seccp2}

We begin by considering the symmetric space sine-Gordon theory resulting from
the Pohlmeyer reduction of the $\mathbb{C}P^2$ $\sigma$-model.
This is a coset $\sigma$-model on $SU(3)/(SU(2) \times U(1))$ which means that
the Lie algebra $\f$ is equal to $ \mathfrak{su}(3)$ while $\f^{(0)} \simeq
\mathfrak{su}(2)
\oplus \mathfrak{u}(1)$.

\paragraph{Automorphism.} Consider therefore $\f= \mathfrak{su}(3)$ with
Chevalley generators $\mathsf{H}_i, \mathsf{E}_i, \mathsf{F}_i$
given in the fundamental representation by
\begin{subequations} \label{sl3 generators}
\begin{alignat}{3}
\mathsf{H}_1 &= \left( \begin{array}{ccc} 1 & 0 & 0\\ 0 & -1 & 0\\ 0 & 0 & 0
\end{array} \right), &\qquad
\mathsf{E}_1 &= \left( \begin{array}{ccc} 0 & 1 & 0\\ 0 & 0 & 0\\ 0 & 0 & 0
\end{array} \right), &\qquad
\mathsf{F}_1 &= \left( \begin{array}{ccc} 0 & 0 & 0\\ 1 & 0 & 0\\ 0 & 0 & 0
\end{array} \right),\\
\mathsf{H}_2 &= \left( \begin{array}{ccc} 0 & 0 & 0\\ 0 & 1 & 0\\ 0 & 0 & -1
\end{array} \right), &\qquad
\mathsf{E}_2 &= \left( \begin{array}{ccc} 0 & 0 & 0\\ 0 & 0 & 1\\ 0 & 0 & 0
\end{array} \right), &\qquad
\mathsf{F}_2 &= \left( \begin{array}{ccc} 0 & 0 & 0\\ 0 & 0 & 0\\ 0 & 1 & 0
\end{array} \right)
\end{alignat}
\end{subequations}
and let $\mathsf{E}_3 = [\mathsf{E}_1, \mathsf{E}_2]$ and $\mathsf{F}_3 = [\mathsf{F}_2, \mathsf{F}_1]$.
The $\mathbb{Z}_2$-automorphism $\sigma$ of $\f$ is defined by
\begin{alignat*}{4}
\sigma(\mathsf{H}_1) &= \mathsf{H}_1, &\qquad
\sigma(\mathsf{H}_2) &= \mathsf{H}_2, &\qquad
\sigma(\mathsf{E}_1) &= \mathsf{E}_1, &\qquad
\sigma(\mathsf{F}_1) &= \mathsf{F}_1,\\
\sigma(\mathsf{E}_2) &= - \mathsf{E}_2, &\qquad
\sigma(\mathsf{F}_2) &= - \mathsf{F}_2, &\qquad
\sigma(\mathsf{E}_3) &= - \mathsf{E}_3, &\qquad
\sigma(\mathsf{F}_3) &= - \mathsf{F}_3.
\end{alignat*}
Note that this is an inner automorphism since $\sigma(x) = g x g^{-1}$ where $g
= \Diag(1, 1, -1)$. We take the bases for the corresponding eigenspaces
$\f^{(0)}, \f^{(1)} \subset \f$ of eigenvalue $\pm 1$ to be
\begin{align*}
\f^{(0)} = \langle \mathsf{H}_1, \mathsf{E}_1, \mathsf{F}_1, \mathsf{H}'_2
= \mathsf{H}_2 + \ha \mathsf{H}_1 \rangle, \qquad
\f^{(1)} = \langle \mathsf{E}_2, \mathsf{F}_2, \mathsf{E}_3, \mathsf{F}_3
\rangle.
\end{align*}
Notice that $\mathsf{H}'_2$ commutes with $\mathsf{H}_1, \mathsf{E}_1,
\mathsf{F}_1 \in \f^{(0)}$ and hence $\f^{(0)} \simeq \mathfrak{su}(2)
\oplus \mathfrak{u}(1)$ as desired.

\paragraph{Casimir decomposition.}
The tensor Casimir of $\mathfrak{su}(3)$ reads
\begin{equation} \label{Casimir su3}
C = \mbox{\small $\frac{1}{6}$} \big( \mathsf{H}_1 \otimes \mathsf{H}_2 +
\mathsf{H}_2 \otimes \mathsf{H}_1 \big) + \mbox{\small $\frac{1}{3}$}
\big( \mathsf{H}_1 \otimes \mathsf{H}_1 + \mathsf{H}_2 \otimes \mathsf{H}_2
\big)
+ \mbox{\small $\frac{1}{2}$} \textstyle\sum_{i = 1}^3 \big( \mathsf{E}_i
\otimes
\mathsf{F}_i + \mathsf{F}_i \otimes \mathsf{E}_i \big),
\end{equation}
and decomposes as $C = C^{(00)} + C^{(11)}$ with respect to the subspaces $\f =
\f^{(0)} \oplus \f^{(1)}$ where
\begin{align*}
C^{(00)} &= \mbox{\small $\frac{1}{3}$} \mathsf{H}'_2 \otimes \mathsf{H}'_2 +
\qa \mathsf{H}_1 \otimes \mathsf{H}_1 + \ha \big( \mathsf{E}_1 \otimes
\mathsf{F}_1 + \mathsf{F}_1 \otimes \mathsf{E}_1 \big),\\
C^{(11)} &= \mbox{\small $\frac{1}{2}$} \big( \mathsf{E}_2 \otimes \mathsf{F}_2
+ \mathsf{F}_2 \otimes \mathsf{E}_2 + \mathsf{E}_3 \otimes \mathsf{F}_3 +
\mathsf{F}_3 \otimes \mathsf{E}_3 \big).
\end{align*}

\paragraph{Classical $r$-matrices.}

For the skew-symmetric solution $\alpha$ of the mCYBE on $\g = \f^{(0)}$
we shall take the standard solution on its $\mathfrak{su}(2)$ part, namely
\begin{equation*}
\alpha = \ha \big( \mathsf{E}_1 \otimes \mathsf{F}_1 - \mathsf{F}_1 \otimes
\mathsf{E}_1 \big).
\end{equation*}
The corresponding non-skew-symmetric solutions $b$ and $c$ of the CYBE read
\begin{subequations} \label{bc CP2}
\begin{align}
b = - \mbox{\small $\frac{1}{3}$} \mathsf{H}'_2 \otimes \mathsf{H}'_2 - \qa
\mathsf{H}_1 \otimes \mathsf{H}_1 - \mathsf{E}_1 \otimes \mathsf{F}_1,\\
c = - \mbox{\small $\frac{1}{3}$} \mathsf{H}'_2 \otimes \mathsf{H}'_2 - \qa
\mathsf{H}_1 \otimes \mathsf{H}_1 - \mathsf{F}_1 \otimes \mathsf{E}_1.
\end{align}
\end{subequations}
The spectral parameter dependent $r$-matrices $a(\lambda)$ and $d(\lambda)$ may
then be written as
\begin{subequations} \label{ad CP2}
\begin{align}
a(\lambda) &= - \frac{1}{1 - \lambda^2} \, b - \frac{\lambda^2}{1 - \lambda^2}
\, c + \frac{2 \lambda}{1 - \lambda^2} C^{(11)}, \\
d(\lambda) &= - \frac{1}{1 - \lambda^2} \, c - \frac{\lambda^2}{1 - \lambda^2}
\, b + \frac{2 \lambda}{1 - \lambda^2} C^{(11)}.
\end{align}
\end{subequations}

\paragraph{Quantum $R$-matrices.}
One can check that solutions of the QYBE with classical limits \eqref{bc CP2} as
$\hbar \to 0$ with $q = e^{i \hbar}$, are given respectively by
\begin{align*}
\mathcal{B}(q) &= q^{- \mbox{\small $\frac{1}{3}$} \mathsf{H}'_2 \otimes
\mathsf{H}'_2 - \qa \mathsf{H}_1 \otimes \mathsf{H}_1} + q^{- \mbox{\small
$\frac{1}{3}$}}  (1- q) \mathsf{E}_1 \otimes \mathsf{F}_1,\\
\mathcal{C}(q) &= q^{- \mbox{\small $\frac{1}{3}$} \mathsf{H}'_2 \otimes
\mathsf{H}'_2 - \qa \mathsf{H}_1 \otimes \mathsf{H}_1} + q^{- \mbox{\small
$\frac{1}{3}$}}  (1- q) \mathsf{F}_1 \otimes \mathsf{E}_1.
\end{align*}
Quantizations of the matrices \eqref{ad CP2} then take the following form  
\begin{subequations} \label{AD CP2}
\begin{align}
\label{A CP2}
\mathcal{A}(q, \lambda) &= \frac{q^{\frac{1}{3}}}{q^{\frac{1}{3}} - \lambda^2}
\,
\mathcal{B}(q)^{-1} - \frac{\lambda^2}{q^{\frac{1}{3}} - \lambda^2} \,
\mathcal{C}(q)
+ \frac{2 q^{-\frac{1}{3}} (q - 1) \lambda}{q^{\frac{1}{3}} - \lambda^2} \,
C^{(11)},\\
\label{D CP2}
\mathcal{D}(q, \lambda) &= \frac{q^{\frac{1}{3}}}{q^{\frac{1}{3}} - \lambda^2}
\, \mathcal{C}(q)^{-1} - \frac{\lambda^2}{q^{\frac{1}{3}} - \lambda^2} \,
\mathcal{B}(q) + \frac{2 q^{-\frac{1}{3}} (q - 1) \lambda}{q^{\frac{1}{3}} -
\lambda^2} \, C^{(11)}.
\end{align}
\end{subequations}

\paragraph{Properties.}
Aside from the general properties listed in section \ref{quantumlattice}, the
quantum $R$-matrices just defined also satisfy $\mathcal{D}_{\1\2}(q, \lambda) =
\mathcal{A}_{\2\1}(q, \lambda)$. The unitarity property \eqref{unitAD}
explicitly reads
\begin{equation*}
\mathcal{A}_{\1\2}(q, \lambda) \mathcal{A}_{\2\1}(q, \lambda^{-1}) =
\frac{(q-\lambda ^2) (q^{-1} -  \lambda^2)}{(q^{\frac{1}{3}} - \lambda ^2)
(q^{-\frac{1}{3}}  - \lambda^2)} \mathsf{1},
\end{equation*}
while the relations \eqref{local to global} and \eqref{quantum integrability
property} hold
for $\gamma(q)=\mathsf{1}$ and $\widetilde{\gamma}(q) = \Diag(q,1,1)$.

\paragraph{Connection with universal $R$-matrix.}
The $R$-matrix in \eqref{A CP2} turns out to be related to the untwisted affine Lie algebra 
$A^{(1)}_2$, the $R$-matrix of which in the fundamental representation was 
given in \cite{Babelon:1981kc}. In order to see this connection explicitly,
it is convenient to use the results
of \cite{Boos:2010ss}, where the universal $R$-matrix obtained
by Khoroshkin and Tolstoy in \cite{tolstoy_1992,Khoroshkin_1992,Khoroshkin_1991} for
$A_2^{(1)}$ was evaluated in the fundamental representation.
One can directly check that the $R$-matrix \eqref{A CP2} obtained here is
proportional to $R^{(2,0,1)}$ in the notation of \cite{Boos:2010ss} with the
replacement $q \to q^{\frac{1}{2}}$. This connection with the untwisted affine Lie algebra 
$A^{(1)}_2$ stems from the automorphism $\sigma$ being inner. In fact, the twisting by the inner automorphism $\sigma$ can be undone at the
quantum level by considering 
\beqz
\widehat{\mathcal{A}}_{\1\2}(q, \lambda/\mu)  = g_{\1}(\lambda) g_{\2}(\mu)
\mathcal{A}_{\1\2}(q, \lambda/\mu) g_{\1}(\lambda)^{-1} g_{\2}(\mu)^{-1}
\eeqz
where $g(\lambda)$ is the diagonal matrix defined as
$g(\lambda) = \Diag(1,1,\lambda)$. Up to some overall scalar factor and the
replacement $q \to q^{\frac{1}{2}}$, this is precisely the $R$-matrix
$R^{(2,0,0)}$ in the notation of \cite{Boos:2010ss}.

\subsubsection{$SU(3)/SO(3)$ symmetric space sine-Gordon model}
\label{secsu3so3}

\paragraph{Automorphism.}

We use the same notations \eqref{sl3 generators} as in section \ref{seccp2}
for the generators in the fundamental representation of $\f =\mathfrak{su}(3)$.
In the case at hand, the $\mathbb{Z}_2$-automorphism acts on an element $x$ of
$\f$ as
\begin{equation} \label{sigmasu3so3}
\sigma(x) = - \eta x^t \eta^{-1},
\end{equation}
where the pseudo-metric $\eta$ is defined by
\beq
\eta=\left(
\begin{array}{ccc}
0&0&1\cr
0&-1&0\cr
1&0&0
\end{array}
\right). \label{eta}
\eeq
The  corresponding eigenspaces $\f^{(0)}$ and $\f^{(1)}$ of $\sigma$
are generated by
\beqz
\f^{(0)} = \langle  \mathsf{h}^{(0)}, \mathsf{e}^{(0)},  \mathsf{f}^{(0)}
\rangle,
\qquad  \f^{(1)} = \langle  \mathsf{h}^{(1)}, \mathsf{e}^{(1)},
\mathsf{f}^{(1)},
\mathsf{E}_3, \mathsf{F}_3 \rangle,
\eeqz
where we have introduced the following linear combinations of the generators
\beqz
\begin{split}
\mathsf{h}^{(0)} = \mathsf{H}_1 + \mathsf{H}_2, \qquad
\mathsf{e}^{(0)} &= \mathsf{E}_1 + \mathsf{E}_2, \qquad
\mathsf{f}^{(0)} = \mathsf{F}_1 + \mathsf{F}_2, \\
\mathsf{h}^{(1)} = \mathsf{H}_1 - \mathsf{H}_2, \qquad
\mathsf{e}^{(1)} &= \mathsf{E}_1 - \mathsf{E}_2, \qquad
\mathsf{f}^{(1)} = \mathsf{F}_1 - \mathsf{F}_2.
\end{split}
\eeqz
Note in particular that $\f^{(0)} \simeq \mathfrak{so}(3)$.

\paragraph{Casimir decomposition.}
The tensor Casimir \eqref{Casimir su3} of $\mathfrak{su}(3)$ decomposes with
respect to the above $\mathbb{Z}_2$-grading as
\begin{align*}
C^{(00)} &= \qa \mathsf{h}^{(0)} \otimes \mathsf{h}^{(0)} +
\qa \mathsf{e}^{(0)} \otimes \mathsf{f}^{(0)} +
\qa \mathsf{f}^{(0)} \otimes \mathsf{e}^{(0)},\\
C^{(11)} & = {\mbox{\small $\frac{1}{12}$}}  \mathsf{h}^{(1)} \otimes
\mathsf{h}^{(1)}
+ \qa  \mathsf{e}^{(1)} \otimes \mathsf{f}^{(1)}
+ \qa  \mathsf{f}^{(1)} \otimes \mathsf{e}^{(1)}
+ \ha \left( \mathsf{E}_3 \otimes \mathsf{F}_3 + \mathsf{F}_3 \otimes
\mathsf{E}_3 \right).
\end{align*}

\paragraph{Classical $r$-matrices.}

Our choice for $\alpha$ is again the standard skew-symmetric solution of the
mCYBE on $\f^{(0)}$, namely
\begin{equation*}
\alpha = \qa \big(  \mathsf{e}^{(0)} \otimes \mathsf{f}^{(0)} -
 \mathsf{f}^{(0)} \otimes \mathsf{e}^{(0)}\big).
\end{equation*}
The corresponding non-skew-symmetric solutions $b$ and $c$ of the CYBE  take the
following form
\begin{subequations} \label{su3so3 b c}
\begin{align}
b &= - \qa \mathsf{h}^{(0)} \otimes \mathsf{h}^{(0)} - \ha \mathsf{e}^{(0)}
\otimes \mathsf{f}^{(0)},\\
c &= - \qa \mathsf{h}^{(0)} \otimes \mathsf{h}^{(0)} - \ha \mathsf{f}^{(0)}
\otimes \mathsf{e}^{(0)}.
\end{align}
\end{subequations}
In terms of these, the $r$-matrices $a(\lambda)$ and $d(\lambda)$
are given by the same expressions as in \eqref{ad CP2}.

\paragraph{Quantum $R$-matrices.}
Quantizations of the classical $r$-matrices \eqref{su3so3 b c}
are given by
\begin{subequations} \label{su3so3 B C}
\begin{align}
\mathcal{B}(q) &= q^{-\frac{1}{4} \mathsf{h}^{(0)} \otimes \mathsf{h}^{(0)} }
\Big( \mathsf{1} - \big( q^{\frac{1}{4}} - q^{- \frac{1}{4}} \big)
\mathsf{e}^{(0)} \otimes \mathsf{f}^{(0)}
+ \big(1 - q^{- \frac{1}{4}}\big)\big(q^{\frac{1}{4}} - q^{- \frac{1}{4}}\big) (
\mathsf{e}^{(0)})^2 \otimes
( \mathsf{f}^{(0)})^2
\Big),\\
\mathcal{C}(q) &= q^{-\frac{1}{4} \mathsf{h}^{(0)} \otimes \mathsf{h}^{(0)} }
\Big( \mathsf{1} - \big( q^{\frac{1}{4}} - q^{- \frac{1}{4}} \big)
\mathsf{f}^{(0)} \otimes \mathsf{e}^{(0)}
+ \big(1 - q^{- \frac{1}{4}}\big)\big(q^{\frac{1}{4}} - q^{- \frac{1}{4}}\big) (
\mathsf{f}^{(0)})^2 \otimes
( \mathsf{e}^{(0)})^2
\Big).
\end{align}
\end{subequations}
As for the $R$-matrix $\mathcal{A}(q,\lambda)$,
based on the established connection of the previous example with
the untwisted affine Lie algebra $A_2^{(1)}$, it is natural to expect a similar relation
in the present case but this time with the twisted affine Lie algebra
$A^{(2)}_2$. The $R$-matrix of the latter in the fundamental
representation was computed in \cite{Izergin:1980pe}. For our purposes
we shall use the results of \cite{Boos:2011zi} in which
a family of $R$-matrices in the fundamental representation
of $\mathfrak{su}(3)$ parametrized by two integers $s_0$ and
$s_1$ was obtained from the universal $R$-matrix \cite{tolstoy_1992,Khoroshkin_1992,Khoroshkin_1991}. Specifically, we will
construct $\mathcal{A}(q,\lambda)$ from the particular solution with $s_0=1$
and $s_1=0$ which can be rewritten as
follows. Introduce a $q$-deformation $\eta_q$ of the metric \eqref{eta} as
\begin{equation*}
\eta_q=\left(
\begin{array}{ccc}
0&0&q^{\frac{1}{4}}\cr
0&-1&0\cr
1&0&0
\end{array}
\right).
\end{equation*}
We use this to define the following $q$-deformation of the automorphism
\eqref{sigmasu3so3}
\begin{equation*}
\sigma_q(x)=-\eta_q x^t \eta_q^{-1}.
\end{equation*}
Then, up to some overall factor and the replacement $q \to q^{1/4}$, the
$R$-matrix considered in \cite{Boos:2011zi}
may be rewritten as
\begin{equation} \label{su3so3 A hat}
\widehat{\mathcal{A}}(q,\lambda) =
\frac{q^{\frac{1}{4}}}{q^{\frac{1}{2}}-\lambda}\mathcal{B}(q)^{-1}
- \frac{q^{\frac{1}{4}} \lambda}{q^{\frac{1}{2}}-\lambda}\mathcal{C}(q)
- \frac{\lambda (q^{\frac{1}{2}}-1)
\big(1+q^{\frac{3}{4}}\big)}{(q^{\frac{1}{2}} - \lambda )\big(\lambda +
q^{\frac{3}{4}}\big)}
\sum_{i=1}^{3}\sum_{j=1}^{3}E_{ij}\otimes \sigma_q(E_{ji}),
\end{equation}
where $E_{ij}$ denotes the $3 \times 3$
matrix whose only non-zero entry is a 1 in the $i^{\rm th}$ row and $j^{\rm th}$
column.
 The desired quantum $R$-matrix with the correct classical limit $a(\lambda)$
given in
\eqref{ad CP2} may now be obtained by rescaling \eqref{su3so3 A hat} as
\begin{subequations} \label{su3so3 A D}
\begin{equation} \label{su3so3 A}
\mathcal{A}(q,\lambda) =  \frac{(q^{\frac{1}{2}} - \lambda )
\big(q^{\frac{3}{4}}+\lambda
\big)}{q^{\frac{1}{4}}\big(q^{\frac{1}{6}}- \lambda \big) \big(q^{\frac{7}{12}}+
 \lambda \big)} \widehat{\mathcal{A}}(q,\lambda).
\end{equation}
Finally, the matrix $\mathcal{D}(q,\lambda)$ is defined through the relation
\eqref{AB=CD}. Such a definition
automatically satisfies the classical limit \eqref{CD classical limit} and is
given explicitly by
\begin{equation} \label{su3so3 D}
\mathcal{D}(q,\lambda) =  \frac{(q^{\frac{1}{2}} - \lambda )
\big(q^{\frac{3}{4}}+\lambda
\big)}{q^{\frac{1}{4}}\big(q^{\frac{1}{6}}- \lambda \big) \big(q^{\frac{7}{12}}+
 \lambda \big)} \widehat{\mathcal{D}}(q,\lambda),
\end{equation}
\end{subequations}
where the quantum $R$-matrix $\widehat{\mathcal{D}}(q, \lambda)$ admits a
similar expression to \eqref{su3so3 A hat}, namely
\begin{equation} \label{su3so3 D hat}
\widehat{\mathcal{D}}(q,\lambda) =
\frac{q^{\frac{1}{4}}}{q^{\frac{1}{2}} - \lambda}\mathcal{C}(q)^{-1}
- \frac{q^{\frac{1}{4}} \lambda}{q^{\frac{1}{2}} - \lambda}\mathcal{B}(q)
- \frac{\lambda (q^{\frac{1}{2}}-1)
\big(1+q^{\frac{3}{4}}\big)}{(q^{\frac{1}{2}} - \lambda )\big(\lambda
+q^{\frac{3}{4}}\big)}
\sum_{i=1}^{3}\sum_{j=1}^{3} \sigma_q(E_{ij}) \otimes E_{ji}.
\end{equation}

\paragraph{Properties.}

The matrices \eqref{su3so3 B C} and \eqref{su3so3 A D} so defined satisfy all
the properties discussed
in section \ref{quantumlattice} as well as the further property
$\mathcal{D}_{\1\2}(q, \lambda) = \mathcal{A}_{\2\1}(q, \lambda)$. Furthermore,
the rescaled $R$-matrices \eqref{su3so3 A hat} and \eqref{su3so3 D hat} are both
unitary, namely
\begin{equation*}
\widehat{\mathcal{A}}_{\1\2}(q,\lambda)
\widehat{\mathcal{A}}_{\2\1}(q,\lambda^{-1}) = \mathsf{1}, \qquad
\widehat{\mathcal{D}}_{\1\2}(q,\lambda)
\widehat{\mathcal{D}}_{\2\1}(q,\lambda^{-1}) = \mathsf{1}.
\end{equation*}
Finally, the relations \eqref{local to global} and \eqref{quantum integrability
property} are
satisfied with $\gamma(q) = \mathsf{1}$ and $
\widetilde{\gamma}(q) =\Diag(q^{\frac{1}{4}} ,1,q^{-\frac{1}{4}})$.

\section{Affine quantum braided group for $\mathfrak{gl}(4|4)$}
\label{secgl44}

Throughout this section we take $\f = \mathfrak{gl}(4|4)$, a basis of which in
the fundamental
representation is given by the $8 \times 8$ matrices $E_{i,j}$ whose only
non-zero entry is a $1$
in the $i^{\rm th}$ row and $j^{\rm th}$ column.

\paragraph{Automorphism.}
The $\mathbb{Z}_4$-automorphism $\sigma$ of $\f$ with the property $\sigma^4 =
\text{id}$ is defined by
\begin{equation*}
\sigma(x) = - K x^{st} K^{-1},
\end{equation*}
where $x^{st}$ denotes the usual supertranspose of the matrix $x$
and $K = \mathsf{1}_4 \otimes i \sigma_2$.
The projection $p^{(k)}$ of $\f$ onto the corresponding eigenspace $\f^{(k)}$ of
$\sigma$ is defined for any $x \in \f$ by
\begin{equation}
p^{(k)}(x) = \qa \bigl(x + i^{3k} \sigma(x) + i^{2k} \sigma^2(x) + i^k
\sigma^3(x)
\bigr). \label{defofproj}
\end{equation}
The subalgebra $\f^{(0)}$ corresponds to two copies of the Lie algebra $\mathfrak{so}(5)$ and
is spanned by
\begin{equation*}
\f^{(0)} = \big\langle \{ \mathsf{h}_i^{(0)} \}_{i=1}^4, \{
\mathsf{e}_i^{(0)}\}_{i=1}^8, \{ \mathsf{f}_i^{(0)} \}_{i=1}^8 \big\rangle,
\end{equation*}
where the basis vectors are given explicitly in terms of the $E_{i,j}$ as
\begin{gather*}
\mathsf{h}_1^{(0)} = E_{1,1}-E_{2,2}, \quad
\mathsf{h}_2^{(0)} = E_{3,3}-E_{4,4}, \quad
\mathsf{h}_3^{(0)} = E_{5,5}-E_{6,6}, \quad
\mathsf{h}_4^{(0)} = E_{7,7}-E_{8,8}, \quad \\
\mathsf{e}_1^{(0)} = E_{3,4}, \quad
\mathsf{e}_2^{(0)} = \frac{E_{4,2}-E_{1,3}}{\sqrt{2}}, \quad
\mathsf{e}_3^{(0)} = \frac{E_{1,4}+E_{3,2}}{\sqrt{2}}, \quad
\mathsf{e}_4^{(0)} = E_{1,2},\\
\mathsf{e}_5^{(0)} = E_{8,7}, \quad
\mathsf{e}_6^{(0)} = \frac{E_{7,5}-E_{6,8}}{\sqrt{2}}, \quad
\mathsf{e}_7^{(0)} = \frac{E_{6,7}+E_{8,5}}{\sqrt{2}}, \quad
\mathsf{e}_8^{(0)} = E_{6,5},\\
\mathsf{f}_i^{(0)} = \big( \mathsf{e}_i^{(0)} \big)^t.
\end{gather*}

\paragraph{Casimir decomposition.}

The tensor Casimir $C$ takes the simple form
\begin{equation*}
C= \sum_{i,j = 1}^8 E_{i,j} \otimes W E_{j,i}
\end{equation*}
where we have introduced the diagonal matrix $W = \Diag
(\mathsf{1}_4,-\mathsf{1}_4)$.
The four components $C^{(00)}$, $C^{(22)}$, $C^{(13)}$ and $C^{(31)}$
of $C$ are obtained by applying the appropriate projections in
\eqref{defofproj}.
For $C^{(00)}$ we find
\begin{equation} \label{C00 gl44}
C^{(00)} = \ha \sum_{i=1}^4 \mathsf{h}_i^{(0)} \otimes W \mathsf{h}_i^{(0)} +
\sum_{i=1}^8 \bigl(\mathsf{e}_i^{(0)} \otimes W \mathsf{f}_i^{(0)} +
\mathsf{f}_i^{(0)} \otimes W \mathsf{e}_i^{(0)} \bigr).
\end{equation}

\paragraph{Classical $r$-matrices.}
Based on the form \eqref{C00 gl44} of the Casimir component $C^{(00)}$, we make
the following choice for the matrix $\alpha$,
\begin{equation*}
\alpha =  \sum_{i=1}^8 \bigl(\mathsf{e}_i^{(0)} \otimes W \mathsf{f}_i^{(0)} -
\mathsf{f}_i^{(0)} \otimes W \mathsf{e}_i^{(0)} \bigr).
\end{equation*}
It is straightforward to check that this is a solution of the mCYBE on $\f^{(0)}$ and is
skew-symmetric. The corresponding non-skew-symmetric constant solutions of the
CYBE read
\begin{subequations} \label{bandcgl44}
 \begin{align} \label{gl44 b}
b &=  -\ha \sum_{i=1}^4 \mathsf{h}_i^{(0)} \otimes W \mathsf{h}_i^{(0)} - 2
 \sum_{i=1}^8 \mathsf{e}_i^{(0)} \otimes W \mathsf{f}_i^{(0)},\\
c &= -\ha \sum_{i=1}^4 \mathsf{h}_i^{(0)} \otimes W \mathsf{h}_i^{(0)} - 2
 \sum_{i=1}^8 \mathsf{f}_i^{(0)} \otimes W \mathsf{e}_i^{(0)}.
\end{align}
\end{subequations}
The classical $r$-matrices $a(\lambda)$ and $d(\lambda)$ may then be
written in the form
\begin{subequations} \label{gl44 a d}
\begin{align}
\label{gl44 a}  a(q, \lambda) = - \frac{1}{1+\lambda^2} b + \frac{\lambda^2}
{1+\lambda^2} c + \frac{2 \lambda^2}{1 - \lambda^4} \big( C^{(00)} +
C^{(22)}\big)
+ \frac{2 \lambda}{1- \lambda^4} C^{(13)} + \frac{2 \lambda^3}{1- \lambda^4}
C^{(31)},\\
\label{gl44 d}  d(q, \lambda) = - \frac{1}{1+\lambda^2} c + \frac{\lambda^2}
{1+\lambda^2} b + \frac{2 \lambda^2}{1 - \lambda^4} \big( C^{(00)} +
C^{(22)}\big)
+ \frac{2 \lambda}{1- \lambda^4} C^{(13)} + \frac{2 \lambda^3}{1- \lambda^4}
C^{(31)}.
\end{align}
\end{subequations}

\paragraph{Quantum $R$-matrices}
Quantizations of \eqref{bandcgl44} can be expressed as
\begin{subequations} \label{gl44 B C}
\begin{gather} \label{gl44 B}
\mathcal{B}(q) = q^H E_1(q) E_3(q) E_4(q) E_2(q) E_5(q) E_7(q) E_8(q) E_6(q),\\
\mathcal{C}_{\1\2}(q) = \mathcal{B}_{\2\1}(q), \label{defCgl44}
\end{gather}
\end{subequations}
where the first factor in \eqref{gl44 B} is the $q$-exponential of the Cartan part of
$b$ which reads
\begin{equation*}
H = - \ha \sum_{i = 1}^4 \mathsf{h}^{(0)}_i \otimes W \mathsf{h}^{(0)}_i.
\end{equation*}
The remaining factors in \eqref{gl44 B} are given by $q$-exponentials of
the $q$-analogues of each non-Cartan term in the expression \eqref{gl44 b}
for $b$, namely
\begin{align*}
E_1(q) &= \mathsf{1} \otimes \mathsf{1} + (q^{-1} - q) \, \mathsf{e}^{(0)}_1
\otimes \mathsf{f}^{(0)}_1,\\
E_2(q) &= \mathsf{1} \otimes \mathsf{1} + 2 (q^{-\frac{1}{2}} - q^{\frac{1}{2}})
\, \mathsf{e}^{(0)}_2 \otimes \mathsf{f}^{(0)}_2,\\
E_3(q) &= \mathsf{1} \otimes \mathsf{1} - 2 (q^{-\frac{1}{2}} - q^{\frac{1}{2}})
\, [\mathsf{e}^{(0)}_1, \mathsf{e}^{(0)}_2]_q \otimes [\mathsf{f}^{(0)}_1,
\mathsf{f}^{(0)}_2]_q, \\
E_4(q) &= \mathsf{1} \otimes \mathsf{1} + (q^{-1} - q) \, \big[
[\mathsf{e}^{(0)}_1, \mathsf{e}^{(0)}_2]_q , \mathsf{e}^{(0)}_2 \big]_q \otimes
\big[ [\mathsf{f}^{(0)}_1, \mathsf{f}^{(0)}_2]_q, \mathsf{f}^{(0)}_2 \big]_q ,\\
E_5(q) &= \mathsf{1} \otimes \mathsf{1} + (q - q^{-1}) \, \mathsf{e}^{(0)}_5
\otimes \mathsf{f}^{(0)}_5,\\
E_6(q) &= \mathsf{1} \otimes \mathsf{1} + 2 (q^{\frac{1}{2}} - q^{-\frac{1}{2}})
\, \mathsf{e}^{(0)}_6 \otimes \mathsf{f}^{(0)}_6,\\
E_7(q) &= \mathsf{1} \otimes \mathsf{1} - 2 (q^{\frac{1}{2}} - q^{-\frac{1}{2}})
\, [\mathsf{e}^{(0)}_5, \mathsf{e}^{(0)}_6]_{q^{-1}} \otimes
[\mathsf{f}^{(0)}_5, \mathsf{f}^{(0)}_6]_{q^{-1}}, \\
E_8(q) &= \mathsf{1} \otimes \mathsf{1} + (q - q^{-1}) \, \big[
[\mathsf{e}^{(0)}_5, \mathsf{e}^{(0)}_6]_{q^{-1}} , \mathsf{e}^{(0)}_6
\big]_{q^{-1}} \otimes \big[ [\mathsf{f}^{(0)}_5, \mathsf{f}^{(0)}_6]_{q^{-1}},
\mathsf{f}^{(0)}_6 \big]_{q^{-1}} .
\end{align*}
Here  $[\mathsf{x}, \mathsf{y}]_q = q^{-\frac{1}{2}}
 \mathsf{x} \, \mathsf{y} - q^{\frac{1}{2}} \mathsf{y} \, \mathsf{x}$ denotes
the
$q$-commutator of $\mathsf{x}$ with $\mathsf{y}$.

Quantizations of \eqref{gl44 a d} may now be written in the following form
\begin{subequations} \label{defADgl44}
\begin{align}
\A(q, \lambda) &= \frac{1}{1+\lambda^2} \mathcal{B}(q)^{-1} +
 \frac{\lambda^2}{1+\lambda^2} \C(q) \notag\\
&\qquad\qquad\;\;\, + \big( q^{\frac{1}{2}} - q^{-\frac{1}{2}} \big)
 \left( \frac{2 \lambda^2}{1 - \lambda^4} \big( C^{(00)} + C^{(22)}\big)
+ \frac{2 \lambda}{1- \lambda^4} C^{(13)}_q + \frac{2 \lambda^3}{1- \lambda^4}
C^{(31)}_q \right),\\
\mathcal{D}(q, \lambda) &= \frac{1}{1+\lambda^2} \mathcal{C}(q)^{-1} +
\frac{\lambda^2}{1+\lambda^2} \mathcal{B}(q) \notag\\
&+ \big( q^{\frac{1}{2}} - q^{-\frac{1}{2}} \big) \left(
 \frac{2 \lambda^2}{1 - \lambda^4} \big( C^{(00)} + C^{(22)}\big) +
 \frac{2 \lambda}{1- \lambda^4} q^{-10 H} C^{(13)}_{q^{-1}} +
 \frac{2 \lambda^3}{1- \lambda^4} q^{-10 H} C^{(31)}_{q^{-1}} \right),
\end{align}
\end{subequations}
where $C^{(13)}_q$ and $C^{(31)}_q$ are $q$-deformations of the respective
components $C^{(13)}$ and $C^{(31)}$ of the tensor Casimir. Explicitly, the
$q$-deformation $C^{(13)}_q$ is defined as
\begin{equation*}
C^{(13)}_q = -\frac{1}{2} \sum_{m,n = 1}^4 q^{(\epsilon_n - \epsilon_m)
\overline{H}} \big( E_{m, n+4} - i \sigma( E_{m, n+4} ) \big) \otimes \big(
E_{n+4, m} + i \sigma( E_{n+4, m} ) \big)
\end{equation*}
with $(\epsilon_n)_{n=1}^4 = (0, 4, 1, 3)$, whereas the $q$-deformation
$C^{(31)}_q$ is obtained from this as $C^{(31)}_{q \1\2} = C^{(13)}_{q^{-1}
\2\1}$. Here we have introduced
\begin{equation*}
\overline{H} = \ha \sum_{i = 1}^4  \mathsf{h}^{(0)}_i \otimes
\mathsf{h}^{(0)}_i.
\end{equation*}
We also have the relation $C^{(13)}_q + C^{(31)}_q = C^{(13)} + C^{(31)}$.

\paragraph{Properties}

The matrices \eqref{gl44 B C} and \eqref{defADgl44}
satisfy all the properties listed in section \ref{quantumlattice} with
\beq
\gamma(q) = \Diag(\mathsf{1}_4, q^5 \mathsf{1}_4), \qquad
\widetilde{\gamma}(q) = \Diag(1,q^{-4},q^{-1},q^{-3},1,q^{-4},q^{-1},q^{-3}).
\eeq
Note that the first four and last four entries along the diagonal of
$\widetilde{\gamma}$
are just $q^{-\epsilon_n}$. Concerning the unitarity property we have
\beq
\mathcal{A}_{\1\2}(q,\lambda) \mathcal{A}_{\2\1}(q,\lambda^{-1})=
\frac{\left(q-\lambda^2\right) \left(q^{-1} - \lambda ^2 \right)}{(1 -
\lambda^2)^2} \mathsf{1},
\eeq
and similarly for $\mathcal{D}$.

\section{Conclusion}

We have shown, by way of example, how to quantize the lattice Poisson algebra of
(semi-)symmetric space sine-Gordon models previously identified in
\cite{Delduc:2012qb,Delduc:2012mk,Delduc:2012vq}. The quantum lattice algebras obtained for
the four models considered each provide new interesting examples of the general
formalism laid out in \cite{Freidel:1991jx,Freidel:1991jv}. But moreover, there
is a certain uniformity among these examples which hints at a general framework
for quantizing (semi-)symmetric space sine-Gordon models.

Indeed, in each of the four models considered, the function algebra can be
quantized within the language of affine quantum braided groups. The necessity
for the departure from the conventional set-up of affine quantum groups can be
seen as a remnant of the non-ultralocality of these models at the classical
level. Specifically, the braiding arises as a quantum counterpart of the
regularization prescription \cite{SemenovTianShansky:1995ha} necessary to
unambiguously define the Poisson bracket of the monodromy matrix. This strongly
suggests that the general formalism presented in section \ref{quantumlattice}
should be the appropriate language within which to address the quantization of
(semi-)symmetric space sine-Gordon models.

Furthermore, the examples discussed in section \ref{secA2} indicate a general
procedure for constructing the various $R$-matrices entering the quantized lattice
algebra of these models. Indeed, in the specific cases of the $\mathbb{C}P^2$ and $SU(3)/SO(3)$
symmetric space sine-Gordon models, we have shown how these $R$-matrices can be directly obtained
from the $R$-matrix of, respectively, the untwisted and twisted affine Lie algebras of type
$A_2$ in the fundamental representation through the works of
\cite{Boos:2010ss,Boos:2011zi}.

Finally, in view of ultimately identifying a quantum lattice model for the
theories in question, the next challenge is to find explicit quantum
lattice Lax operators $\hat{\L}^n$ satisfying the algebra given in section
\ref{subsec quant quad algebra}. This is an important problem which we leave for
future work.

\paragraph{Acknowledgements}

We would like to thank D. Fioravanti, 
 J. M. Maillet, H. Samtleben, V. Terras and C. Young for useful discussions.

\providecommand{\href}[2]{#2}\begingroup\raggedright\endgroup

\end{document}